\documentclass{article}
\usepackage{graphicx}
\usepackage{color}
\begin{document}
\title{\bf An Atmospheric Cerenkov Telescope Simulation System}
\author{\bf Srikanta Sinha\\
            3A, Sharda Royale Apt., G. M. Palya,\\
            Bengaluru-560 075, INDIA\\
            e-mail:sinha.srikanta@gmail.com}

\maketitle

\newpage

\section{\bf ABSTRACT}
	A detailed numerical procedure has been developed to simulate the
mechanical configurations and optical properties of
Imaging Atmospheric Cerenkov Telescope systems. To test these
procedures a few existing ACT arrays are simulated. First results from
these simulations are presented.
\section{\bf INTRODUCTION}
         During the last two decades or so the Atmospheric Cerenkov Imaging Technique
(IACT) has developed rapidly, thus opening up 
the very high energy (VHE, 0.1 TeV to 10 TeV) gamma ray astronomical particle band and
has led to the detection and detailed studies of very large number of astrophysical
sources that emit electromagnetic particle radiation in these energies.
Examples of Cerenkov imaging telescopes are (i) the Whipple Observatory, (ii) the MAGIC telescope
system, (ii) the VERITAS telescopes,(iv) the HESS I and HESS II telescopes.
This branch of astronomy is yet to grow further with the commissioning of the Cerenkov
Telescope Array (CTA).

         Before the advent of the Cerenkov imaging technique, several
experimental groups around the world made attempts to detect celestial
VHE gamma ray sources using the traditional (presently somewhat obsolete)
Wavefront Sampling Technique. Unlike the Cerenkov Imaging Technique
that uses stereoscopic imaging of the particle showers initiated by
the celestial VHE gamma rays and also the background cosmic ray
particles, the wavefront sampling technique just attempts to collect
the Cerenkov photons using large mirrors and tries to distinguish
between photon initiated and hadron initiated showers using 
somewhat loosely defined characterisctics, such as the shape of
the lateral distribution of the Cerenkov photons, or the Cerenkov
photon pulse profile etc. This resulted in poor signal-to-noise
ratios.
 
        In the $1980$s a simple computer procedure was developed
at TIFR, Mumbai just to calculate the lateral distribution of the
Cerenkov photons emitted by soft (electro-magnetic) cascades in 
the atmosphere. This Monte Carlo procedure was based on the soft
cascade program developed by (Late) R H Vatcha. The Cerenkov photon
emission routines (mostly analytical) were added to the soft cascade
procedure.

        Later, the Cerenokov emission routines were rewritten
completely to follow individual Cerenkov photons along their
path together with other improvements. 

        Since the Wavefront Sampling is presently almost out
of fashion, one feels the need to develop numerical simulation
procedures for the Cerenkov Imaging Technique. In the present 
we try to give some preliminary results from such an endeavour.

\section{\bf THE DETECTOR SYSTEMS}
	The details of the HAGAR telescope are available elsewhere [SHU].
	This array employs the traditional wave-front sampling
technique. This system consists of seven clusters of mirrors
arranged in the form of a hexagon with side 50 meters. Each
cluster consists of seven mirrors (each having a diameter of 0.9
meters and a focal ratio ($f/d$) equal to 1).

        The details of the HESS I IACT telescope system are given
in [BER03]. Briefly, the HESS I system consists of four individual
imaging telescopes arranged in the form of a square having sides
of 120 meters. Each telescope consists of 382 spherical mirror
facets. All the facets have the same diameter equal to $60 cm$.
The focal length of each facet is $15 m$. The 382 mirror facets
are arranged in the form of a panel with a small central hole.
The HESS I telescopes are located at a site in Namibia at an
altitude of $1800 m$ above the mean sea-level.

   The light reflected by the telescope mirror (having a total
reflector area $ 108 m^{2}$) is focused onto a camera consisting
of 960 pixels. Each pixel consists of a photo-multiplier tube
(PMT, ) having a diameter equal to $        $. A Winston Cone
placed at the face of each PMT facilitates efficient light collection.
The $960$ PMTs are grouped into PMT DRAWERS ($4X4$). Thus, there are
a total of $60$ DRAWERS in the camera. The photo-cathode is a
bi-alkali one with a peak quantum efficiency of $24\%$ at
$340 nm$ wavelength. The camera is placed at a distance of $15 m$
from the reflector, i.e. at its focus.

\section{\bf THE SIMULATION SYSTEM}
             The numerical simulations are carried out using a
hybrid technique consisting of both analytical and Monte Carlo
(based on random numbers). Random numbers are 
used wherever necessary. The simulation codes are written 
largely using the FORTRAN 90/95 language (the
GNU GFORTRAN compiler is used for this purpose).
Details of the simulation procedures are available elsewhere
(the author, in preparation). In the following we give a very
brief outline of the same.

            The present simulation system consists of two major
components. One part of the system deals with the simulation of
the Cerenkov telescope arrays (the complete mechanical, optical
and electronic aspects). This part has been developed recently.
The other part of the simuation code deals with the simulations
of the Extensive Air Showers (EAS) which are the sources that
emit the Cerenkov photons. The EAS are initiated by the Very
High Energy (VHE) gamma rays from the Celestial objects and also
(at the same time much more numerous) Cosmic particle radiation
(mainly protons and other heavier nuclei) that constitute the
unwanted (for this type of work) background.

       For the simulations of Extensive Air Showers (EAS) we do not 
use CORSIKA, but a home-grown software. The electro-magnetic (soft)
cascade program is essentially that developed by R. H. Vatcha [VAT] and
the hadronic part is essentially that developed at TIFR. 
The Cerenkov photon emission procedures
were added by the present author alongwith others at TIFR in the
1980s [BSA82], [RAO88], [SIN95] and later developed further by the present author. 
The main shower simulation procedures have also been significantly
modified. 

\section{\bf RESULTS of SIMULATIONS}
       In Fig.1 we present the simulated profile of a single large 
parabolic mirror (diameter $10 m$).
\begin{figure}
\centering
\includegraphics[width=9cm, height=8cm]{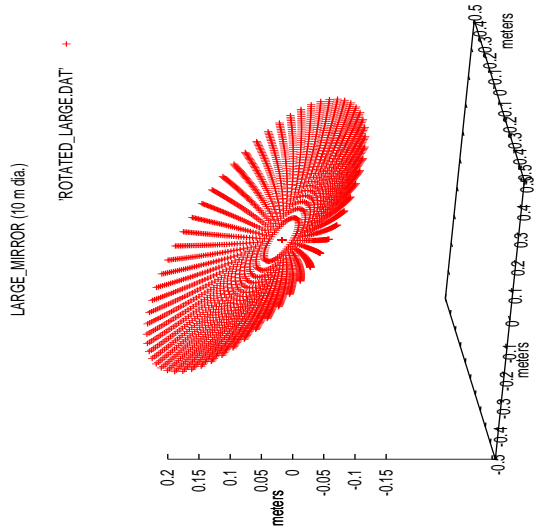}
\caption{A Single Large (10 m) Mirror.}
\end{figure}
       In Fig.2 we present the simulated profile of a single Atmospheric Cerenkov Telescope (this telescope
consists of seven parabolic mirrors, each having a diameter equal to 0.9 meters). This type of telescopes 
are being used in the Indian HAGAR experiment situated at Leh in the western Himalayas.
\begin{figure}
\centering
\includegraphics[width=9cm, height=8cm]{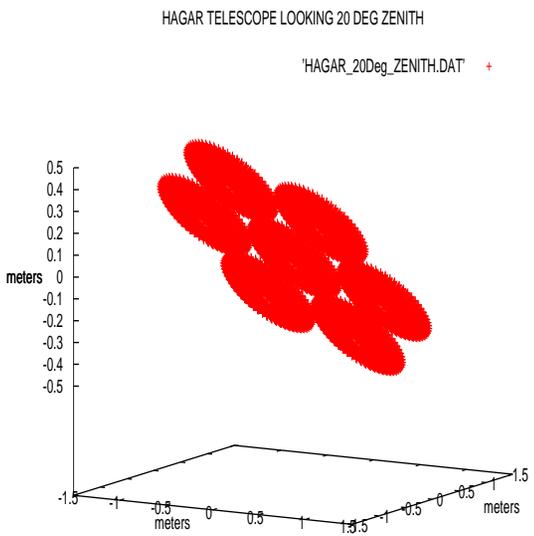}
\caption{The HAGAR telescope looking 20 deg. Zenith.}
\end{figure}
       The HESS I Imaging Atmospheric Telescope System is in operation since in Namibia at an altitude of.
The HESS I System consists of four Imaging telescopes, each one having an effective area of $108 m^{2}$. Each of 
the four HESS I telescope mirror consists of $382$ mirror facets (each having diameter $60 cm$). Each HESS I mirror 
may be thought of as consisting of six large sectors, each sector looking like an inverted triangle.
In Fig.3 we present the simulated image of a single telescope. 
\begin{figure}
\centering
\includegraphics[width=9cm, height=8cm]{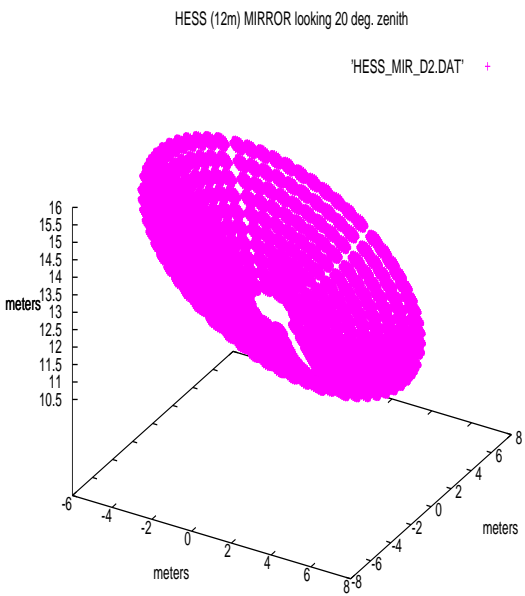}
\caption{One of the four (12 meter)  HESS I Telescopes.}
\end{figure}

      Fig.4 shows the simulated profile of the HESS I camera.
The camera consists of $960$ PMTs.
\begin{figure}
\centering
\includegraphics[width=9cm, height=8cm]{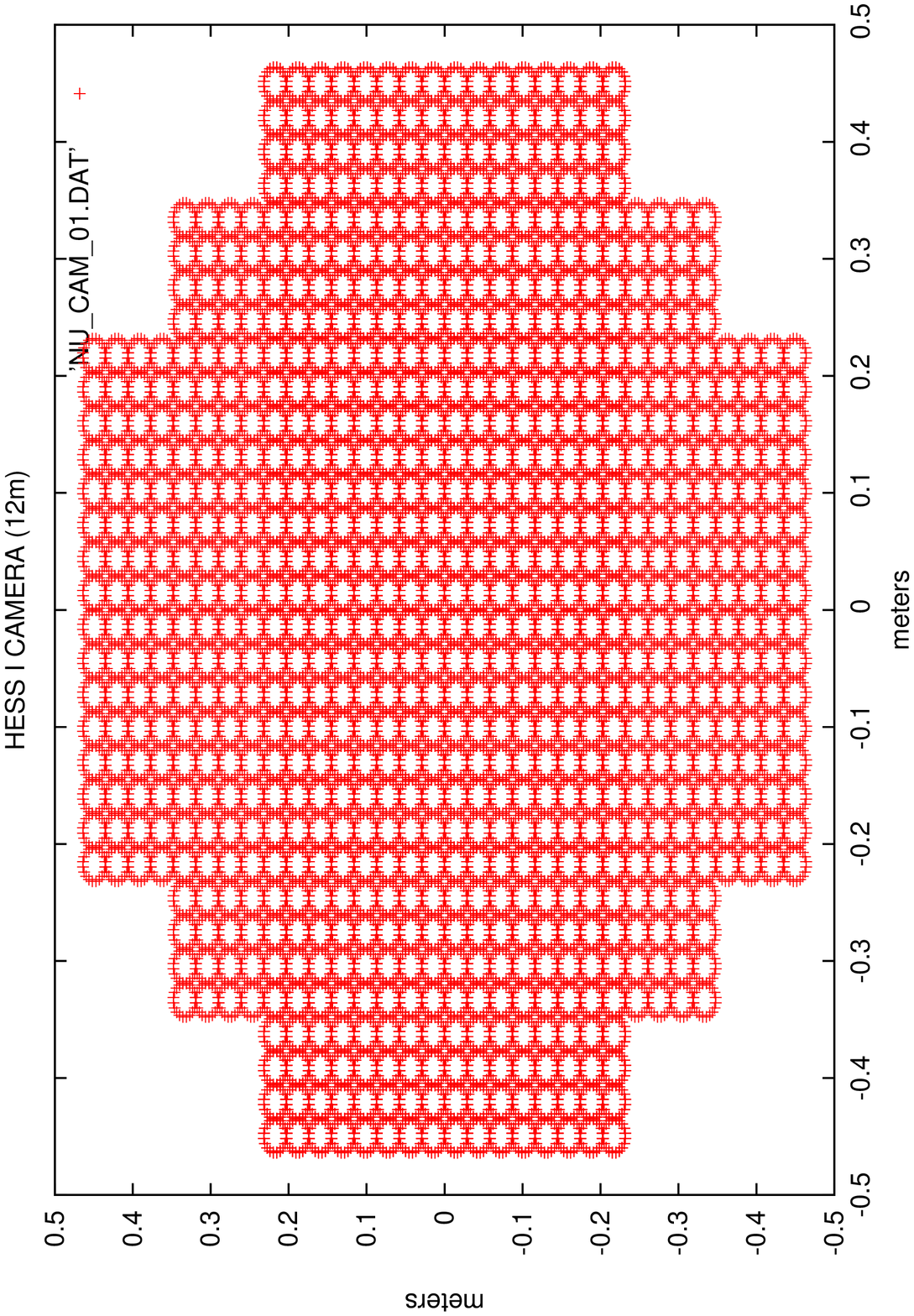}
\caption{The HESS I CAMERA (consisting of 960 PMTs.}
\end{figure}
     Fig.5 shows the camera positioned at the focus of the telescope.
\begin{figure}
\centering
\includegraphics[width=9cm, height=8cm]{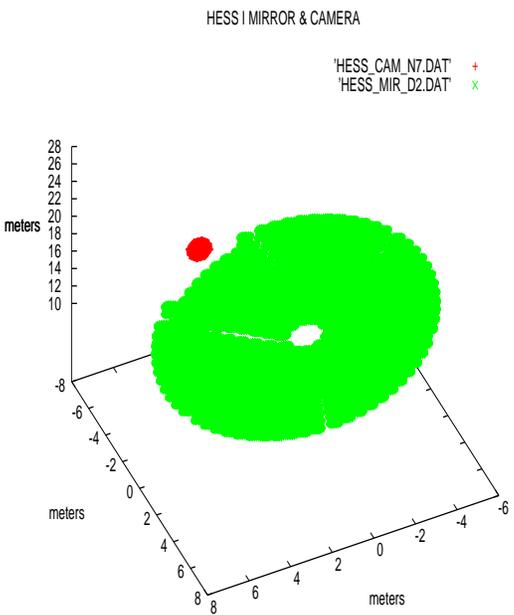}
\caption{The HESS I mirror with the camera at its focus.}
\end{figure}
\section{\bf DISCUSSIONS and CONCLUSION}
    There are certain small inaccuracies in the present 
calculations. These have to be removed. It is possible to
improve the procedures by including finer details.
Hopefully we will be able to present more advanced work in
this direction.

\section{\bf ACKNOWLEDGEMENTS}
    I would like to express my deep sense of gratitude to all my
teachers, especially to (Late) A.N. Ghosh, Prof. A.K. Biswas, 
Dr. B.K. Chatterjee and (Late) K. Sivaprasad.
They had been and remain great sources of inspiration
and support to me. I also
thank the HESS collaboration to make a lot of information available
about this great experiment on their website.

\end{document}